\documentclass[%
 reprint,
superscriptaddress,
 amsmath,amssymb,
 aps,
floatfix,
]{revtex4-1}

\usepackage{graphicx}
\usepackage{dcolumn}
\usepackage{bm}

\usepackage[utf8]{inputenc}
\usepackage[T1]{fontenc}
\usepackage{mathptmx}
\usepackage{times}
\usepackage{xcolor}
\usepackage{siunitx}

\begin{document}

\preprint{AIP/123-QED}
\title[Screened hydrogen model with fractional Coulomb potential]{Generalized scaling law for exciton binding energy in two-dimensional materials}

\author{S. Ahmad}
 \affiliation{ 
Electrical Engineering Department, Information Technology University (ITU) of the Punjab, Lahore 54000, Pakistan
}%
\thanks{The first two authors contributed equally to this work.}
\author{M. Zubair}%
\email{muhammad.zubair@itu.edu.pk}
 \affiliation{ 
NanoTech Lab, Electrical Engineering Department, Information Technology University (ITU) of the Punjab, Lahore 54000, Pakistan
}%
 \thanks{The first two authors contributed equally to this work.}
\author{O. Jalil}%
 \affiliation{ 
Electrical Engineering Department, Information Technology University (ITU) of the Punjab, Lahore 54000, Pakistan
}%

\author{M. Q. Mehmood}%
 \affiliation{ 
NanoTech Lab, Electrical Engineering Department, Information Technology University (ITU) of the Punjab, Lahore 54000, Pakistan
}%

\author{U. Younis}%
 \email{usman.younis@itu.edu.pk}
 \affiliation{ 
Electrical Engineering Department, Information Technology University (ITU) of the Punjab, Lahore 54000, Pakistan
}%
\affiliation{%
College of Materials Science and Engineering, Shenzhen Key Laboratory of Microscale Optical Information Technology, Chinese Engineering and Research Institute of Microelectronics, Shenzhen University, 3688 Nanhai Ave, Shenzhen 518060, People’s Republic of China
}%

\author{X. Liu}
\affiliation{%
College of Materials Science and Engineering, Shenzhen Key Laboratory of Microscale Optical Information Technology, Chinese Engineering and Research Institute of Microelectronics, Shenzhen University, 3688 Nanhai Ave, Shenzhen 518060, People’s Republic of China
}%
\author{K. W. Ang}
\affiliation{%
Department of Electrical and Computer Engineering, National University of Singapore, 4 Engineering Drive 3, Singapore 117583, Singapore
}%
\author{L. K. Ang}
 \email{ricky\textunderscore ang@sutd.edu.sg}
\affiliation{%
Science and Math Cluster, Singapore University of Technology and Design (SUTD), 8 Somapah Road, Singapore 487372, Singapore
}%

\date{\today}
             
\begin{abstract}
 Binding energy calculation in two-dimensional (2D) materials is crucial in determining their electronic and optical properties pertaining to enhanced Coulomb interactions between charge carriers due to quantum confinement and reduced dielectric screening. Based on full solutions of the Schrödinger equation in screened hydrogen model with a modified Coulomb potential ($1/r^{\beta-2}$), we present a generalized and analytical scaling law for exciton binding energy, $E_{\beta} = E_{0}\times \big (\,a\beta^{b}+c\big )\, (\mu/\epsilon^{2})$, where $\beta$ is a fractional-dimension parameter accounted for the reduced dielectric screening. The model is able to provide accurate binding energies, benchmarked with the reported Bethe-Salpeter Equation (BSE) and experimental data, for 58 mono-layer 2D and 8 bulk materials respectively through $\beta$. For a given material, $\beta$ is varied from $\beta$ = 3 for bulk 3D materials to a value lying in the range 2.55$-$2.7 for 2D mono-layer materials. With $\beta_{\text{mean}}$  = 2.625, our model improves the average relative mean square error by 3 times in comparison to existing models. The results can be used for Coulomb engineering of exciton binding energies in the optimal design of 2D materials. 
\end{abstract}
\maketitle

\section{Introduction}
Accurate determination of optical and electronic properties of bulk and 2D materials is crucial for many applications, especially when the excitonic effects become significant in the 2D regime due to structural changes induced by dielectric environments leading to variation in binding energies \cite{grub2015,wu2017excitonic, molina2011phonons, raja2019dielectric,waldecker2019rigid,pandey2019noninvasive,france2017thermal,tanner2018interface,utama2019dielectric,carozo2017optical}. \textcolor{black}{The 2D nature of the material makes the excitons easily tunable, with some external stimuli, enabling excitonic transport based photonic devices such as electrically driven light emitters, opto-valleytronic devices, photovoltaic solar cells and lasers \cite{xiao2017excitons,pospischil2016optoelectronic,butov2017excitonic}.} Thus binding energy calculation in 2D materials has become an active area of research, and numerous studies based on theoretical and experimental approaches have been reported in the literature to understand the excitonic effects and their implications \cite{yip2019tight, park2018direct, steinhoff2015efficient, mueller2018exciton, thygesen2017calculating}. Due to the high cost of experimental procedures, reliance on numerical approaches like Bethe-Salpeter Equation (BSE) becomes inevitable for accurate modeling of excitonic effects and thus serves as a benchmark for other models \cite{haastrup2018computational}. In order to reduce the computational complexity, analytical methods are desirable to speed up the design processes \cite{onida2002electronic,qiu2013optical,lin2014dielectric}. A number of models to describe the exciton behavior are reported, including the pioneer works of Frenkel and Wannier-Mott (W-M) \cite{frenkel1931transformation,wannier1937structure}. The Frenkel model is based on the concept of localized screening, whereas the W-M model incorporates an average delocalized screening effect irrespective of the actual dielectric environment and thus leads to an overestimation of binding energies. Olsen et al. \cite{olsen2016simple} introduced an effective dielectric screening based on material polarizability resulting in relatively accurate binding energies. Recently, a generalized extension of the model by Jiang et al. \cite{PhysRevLett.118.266401} found the 2D exciton binding energy to be one-fourth of the band gap, resulting in error reduction. The difficulty with the existing models is that the inherent Coulomb screening potential, does not take into account the structural confinement effects and therefore the accuracy in 2D binding energy calculations is compromised. Therefore, we propose a simple approach to explicitly incorporate the structural confinement effects in the screening potential by representing the Coulomb potential in a fractional space to represent a more realistic dielectric environment in an effective manner. Moreover, the realistic systems are not 2D in a strict sense, and the fields including electric field, magnetic field, electron emission, angular momentum are not confined to a smooth plane \cite{yang1991analytic}. Hence, the central force between the electron and the nucleus may be better represented by generalized Coulomb potential function, assuming that the system lies in an equivalent fractional-dimensional space. It is worthwhile to mention that the concept of fractional-dimensional space has been successfully applied to study the effects of confinement, roughness, and disorder in various physical problems arising in electron device modeling \cite{zubair2018thickness,zubair2018fractional}, plasma physics \cite{zubair2016fractional} and electromagnetism \cite{zubair2012electromagnetic}. 

Motivated by the above, the purpose of this paper is four-fold: 

First, an analytical model based on the power law fit to the full solution of the simple hydrogen model with a fractional Coulomb potential is presented of the form $E_{\beta} = E_{0}\times (a\beta^{b}+c)(\mu/\epsilon^{2})$, namely fractional Coulomb potential (FCP) model, where $\beta$ is a fractional-dimension parameter linked to the screened Coulomb potential, $E_{0}$ = 13.606 eV is the Rydberg energy, $a = -$2.619$\times$10$^{4}$, $b = -$9.634, $c = -$0.3833, $\mu$ and $\epsilon$ represent material specific reduced exciton effective mass and dielectric constant, respectively.  

Second, we demonstrate the accuracy of the proposed FCP model to calculate the 2D and bulk binding energies, benchmarked with BSE and experimentally reported data. The FCP model allows the incorporation of the structural confinement effects and enables the calculation of exact 2D binding energies corresponding to the actual dielectric environments through fractional-dimension parameter $\beta$ with bulk values intercepted at $\beta = 3$ and mono-layer data represented by $\beta$ lying in the range 2.55$-$2.7. We also show that W-M model is a special case of FCP model reduced at a fixed $\beta = 2.515$. 

Third, we report a correction in the average 2D exciton screening represented at $\beta_{\text{mean}} =  2.625$, based on error analysis of our proposed FCP model in comparison with existing models by using BSE reported data for 58 mono-layer materials having binding energies up to 1.5 eV as reference. It is shown that W-M and Jiang et al. \cite{PhysRevLett.118.266401} based binding energy calculations have average mean square error (MSE) of 39.2$\%$ and 26$\%$, respectively, whereas the proposed FCP model (at $\beta_{\text{mean}} = 2.625$) reduces the average MSE to below 12.8$\%$.  

Finally, we show there exists a simple scaling with a smooth transition of $\beta$ corresponding to structural confinement from bulk to the mono-layer regime. 

\section{Formulation}
The FCP model is based on a fractional Coulomb potential embedded in an infinite quantum well given by the radial part of a simple hydrogen model as:

\begin{equation}
\begin{aligned}
\bigg ( \, -\frac{\hbar^{2}}{2\mu}\frac{d^{2}}{dr^{2}} + V_\text{frac}(r) \bigg )\,\psi(r)=E\psi(r),
\end{aligned}
\label{equation1}
\end{equation}
where $\mu$ = 0.9995$m_{0}$ ($m_{0} = 9.11 \times 10^{-31}$ kg) is the exciton reduced mass, $r$ is the radial distance and $V_{\text{frac}}(r)$ is the fractional Coulomb potential. The fractional Coulomb potential is a generalized form of the standard Coulomb potential which exhibits a Coulomb-like electron-hole pair interaction and is based on the fractional-dimensional Poisson equation~\cite{muslih2007fractional,eid2009fractional,zubair2012electromagnetic} of the form:
\begin{equation}
\begin{aligned}
\nabla_{\beta}^{2}V_{\text{frac}}(r) = -\rho/\epsilon,
\end{aligned}
\label{equation2}
\end{equation}
where $\beta$ is a fractional-dimension parameter valid in the range $2 < \beta \leq 3$, $\rho$ is volume charge density and $\nabla_{\beta}^{2}$ is a generalized fractional-dimensional Laplacian operator in spherical coordinates \cite{stillinger1977axiomatic} with radial component given by:
\begin{equation}
\begin{aligned}
\nabla_{\beta}^{2}=\frac{1}{r^{\beta-1}}\frac{\partial}{\partial r}\bigg[\,r^{\beta-1}\frac{\partial}{\partial r}\bigg].
\end{aligned}
\label{equation3}
\end{equation}

Following the analytical solution of (\ref{equation2}) from~\cite{muslih2007fractional}, the fractional Coulomb potential takes the form:

\begin{equation}
\begin{aligned}
V_\text{frac}(r) = k_{\beta}e^{2}/r^{\beta-2},
\end{aligned}
\label{equation4}
\end{equation}
where $k_{\beta} = \Gamma(\beta/2)/ (2\pi^{\beta/2}(\beta-2)\epsilon_{0})$. At $\beta$ = 3, $k_{\beta}$ reduces to 1/4$\pi\epsilon_{0}$ and $V_\text{frac}(r)$ simplifies to the standard Coulomb potential of the form $1/r$. When $\beta$ differs from three, the Coulomb potential of a point source
falls off as $1/r^{\beta-2}$ and the dynamical symmetry is broken; this is linked to the reduced hydrogenic screening effectively and leads to an effective reduction in the Bohr radius. 

\begin{figure}[ht!]
\begin{centering}
    \includegraphics[width=1.0\columnwidth]{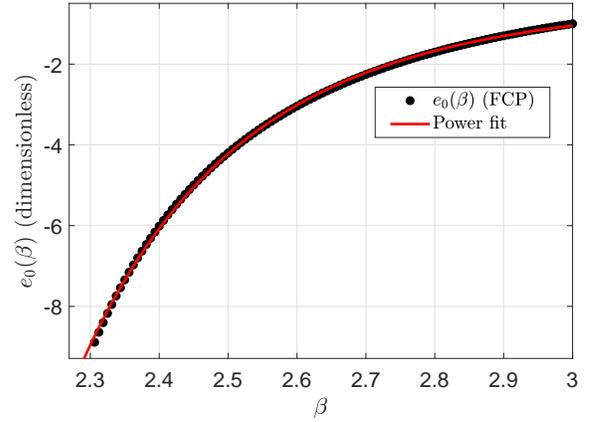}
   
    \caption{Curve fitting to FCP computed ground state hydrogenic energies for 2 $<$ $\beta$ $\leq$ 3, based on full solution. Markers indicate the FCP data points. Dotted line presents power law fit to the data with function $e_{0}(\beta)  = a\beta^{b}+c$ where $a = -$2.619$\times$10$^{4}$, $b = -$9.634 and $c = -$0.3833, with 95$\%$ confidence bound of $(-2.867\times10^{4}, -2.37\times10^{4})$, $(-9.749, -9.519)$ and $(-0.4164, -0.3502)$, respectively.} 
    \label{figure1}
\end{centering}
\end{figure}

The full solution to $(\ref{equation1})$ is performed (see Appendix for detailed derivation) to calculate the ground state hydrogenic energy $e_{0}(\beta)$, where $e_{0}(\beta) = E_{0}(\beta)/13.6$ eV. A power law fit to the full solution of $(\ref{equation1})$ is performed to develop an analytical scaling law for binding energy calculation given by: 

\begin{equation}
\begin{aligned}
E_{\beta} = E_{0}\times \big (\,a\beta^{b}+c\big )\,\bigg (\,\frac{\mu}{\epsilon^{2}}\bigg),
\end{aligned}
\label{equation5}
\end{equation}
where $E_{0} = 13.6$ eV, $a\beta^{b}+c \approx e_{0}(\beta)$, $a = -2.619\times10^{4}$, $b = -9.634$ and $c = -0.3833$ are the fitting parameters obtained from a power law fit to $e_{0}(\beta)$ with 95$\%$ confidence interval of $(-2.867\times10^{4}, -2.37\times10^{4})$, $(-9.749, -9.519)$ and $(-0.4164, -0.3502)$ respectively as shown in Fig. \ref{figure1}. Here $\epsilon$ is the dielectric constant, calculated with $\frac{1}{2}(1+\sqrt{1+\frac{32\pi\alpha\mu}{3}})$, which represents the effective linear screening for strict 2D case \cite{olsen2016simple}, $\mu$ is the reduced exciton mass and $\alpha$ is the 2D polarizability based on G$_{0}$W$_{0}$ reported data \cite{haastrup2018computational}. 

\begin{figure}[ht!]
\begin{centering}
    \includegraphics[width=1.0\columnwidth]{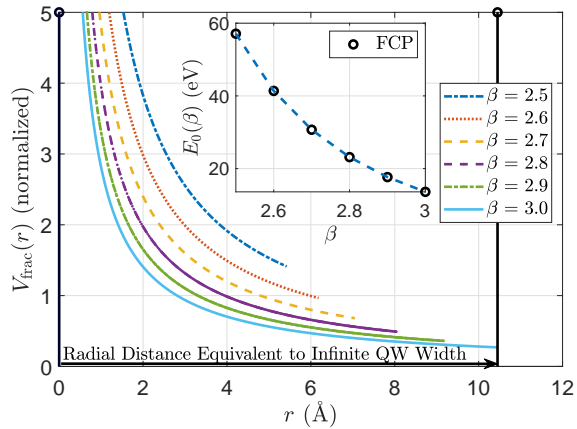}
    \caption{Illustration of the FCP model with an analogous example of an infinite QW model. The radial distance corresponds to the QW width, taken at least 20 times the Bohr radius, in order to converge the full solution of the FCP model in $(\ref{equation1})$ for ground state hydrogenic energies as shown in the inset. The FCP model reduces to a standard hydrogen model at $\beta$ = 3 with $E_{0}(\beta)$ = 13.6 eV, whereas for 2 < $\beta$ < 3, the FCP model reflects a screened hydrogen model and thus results in $E_{0}(\beta)$ > 13.6 eV. The screened hydrogen model corresponds to an infinite QW model, having the provision to vary the QW width through $\beta$.} 
    \label{figure2}
\end{centering}
\end{figure}

\begin{figure}[b]
\begin{centering}
    \includegraphics[width=1.0\columnwidth]{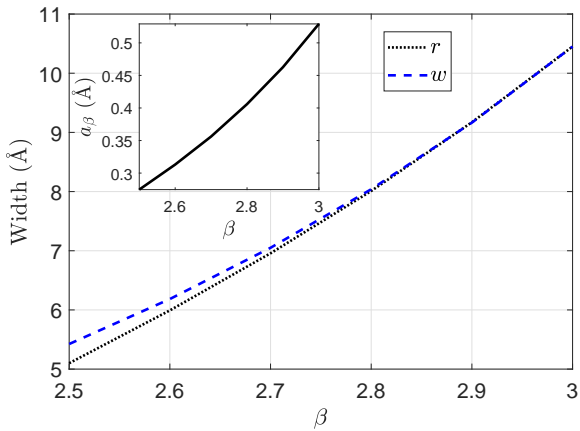}
    \caption{Demonstration of the FCP model as an equivalent infinite QW model. Black dotted line shows the corresponding decrease in the radial distance with $\beta$, as depicted in Fig. \ref{figure2}. Blue dashed line is the equivalent reduction in the infinite QW width calculated using $(\ref{equation6})$, corresponding to ground state hydrogenic energies calculated with the FCP model in $(\ref{equation1})$ for 2.5 $\leq \beta \leq$ 3. The inset shows a reduction in the effective Bohr radius calculated as $a_{\beta} = w/19.75$, where $w$ is the infinite QW width for 2 < $\beta \leq$ 3.} 
    \label{figure3}
\end{centering}
\end{figure}

An infinite quantum well (QW) equivalence of the FCP model in $(\ref{equation1})$ is demonstrated to explain an effective reduction in the Bohr radius. The fractional Coulomb potential for 2.5 $\leq \beta \leq$ 3 is calculated using $(\ref{equation4})$ and is plotted against a radial distance $r$ corresponding to $\beta$ = 3 as illustrated in Fig. \ref{figure2}. A radial distance of approximately 20 times the Bohr radius ($19.75 \times a_{0}$ where $a_{0} = \SI{0.529}{\angstrom}$) with a uniform mesh grid of 300 points is taken to emulate the full wave behaviour of the FCP model, necessary for convergence (see Appendix for details). Here the radial distance is equivalent to an infinite QW width and is highlighted to shrink with decreasing values of $\beta$. This is validated by first, computing the ground state hydrogenic energies $E_{0}(\beta)$ for $2.5 \leq \beta \leq 3$ using the FCP model in $(\ref{equation1})$ as shown in the inset of Fig. \ref{figure2}. The resulting values of $E_{0}(\beta)$ are then employed to calculate the corresponding infinite QW width $w$ using,

\begin{equation}
\begin{aligned}
w = \sqrt[]{\hbar^{2}\pi^{2}/2\mu E_{0}(\beta)},
\end{aligned}
\label{equation6}
\end{equation}

 which is in excellent agreement with the radial distance $r$ as shown in Fig. \ref{figure3}. The decrease in $w$ corresponds to a reduction in the effective Bohr radius given by $a_{\beta} = w/19.75$, as illustrated in inset of Fig. \ref{figure3}. The reduction in $a_{\beta}$ takes place due to a proportionate decrease in the QW dimensions, as a well width of at least 19.75 $\times a_{\beta}$ is required for the infinite QW model to produce hydrogenic energies $E_{0}(\beta)$. 
 
\section{Results and discussion}
\begin{figure}[ht!]
\begin{centering}
    \includegraphics[width=1.0\columnwidth]{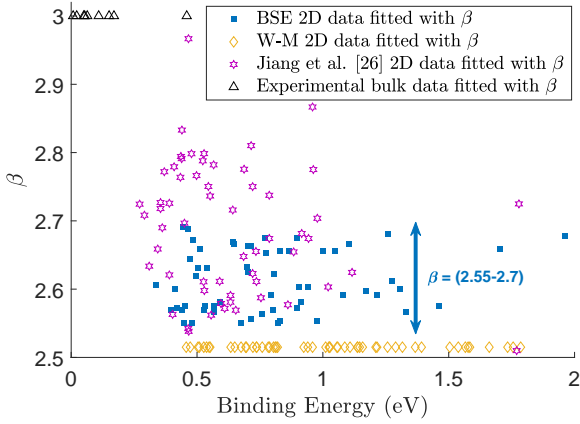}
    \caption{FCP model in $(\ref{equation5})$ based calculations, to intercept $\beta$ for 58 mono-layer materials for reported values of BSE binding energies \cite{haastrup2018computational}, W-M data is calculated for 58 mono-layer materials using $4\times13.6(\frac{\mu}{\epsilon^2})$ and fitted with respective $\beta$ \cite{wannier1937structure,cheiwchanchamnangij2012quasiparticle}, Jiang et al. \cite{PhysRevLett.118.266401} data is calculated for 58 mono-layer materials with $E_{g}/4$ and fitted with respective $\beta$, experimentally reported data for 8 bulk materials is also fitted with respective $\beta$. Here it is to be noted that the binding energy calculation for the aforementioned models has been performed by employing required data reported, the complete details are given in Supplemental Material \cite{SUPPLIMENTAL}. Blue double-arrow shows the range of $\beta$ for 2D BSE data.}
    \label{figure4}
\end{centering}
\end{figure}

 The significance of fractional Coulomb potential in 2D binding energy calculation is highlighted by performing a comparison of the existing analytical methods with the proposed FCP model by using BSE reported binding energies for 58 mono-layers as benchmark \cite{haastrup2018computational}. The analysis is performed to intercept $\beta$ based on the FCP model in $(\ref{equation5})$ for binding energies reported with BSE and that of calculated with existing analytical models. The result in Fig. \ref{figure4} shows the FCP model to accurately produce results through $\beta$ with most of the mono-layer materials found to lie at values of $\beta$ in the range 2.55$-$2.7, to represent the actual 2D exciton screening. Whereas the W-M model is shown to overestimate the 2D binding energies with the average screening represented by a fixed value of $\beta$ = 2.515 as a special case. This is because the W-M model considers the average screening in 2D excitons, with binding energies given by 4$\times$(13.6$\mu$/$\epsilon^{2})$ \cite{yang1991analytic}. However, the exciton model given by Jiang et al. \cite{PhysRevLett.118.266401} results in significant under- or over-estimation of binding energies. Moreover, the FCP model is also demonstrated to intercept binding energies for 8 bulk materials at $\beta$ = 3, benchmarked with experimentally reported data  \cite{fortin1975excitons,laturia2018dielectric,habenicht2018mapping,anedda1980exciton,ugeda2014giant,cao2013two,arora2017interlayer,beal1976excitons,beal1979kramers,museur2011exciton,shan1996binding,asahina1984band}. 
 
 The fractional-dimension parameter $\beta$ is linked to the screening effect due to the fractional Coulomb potential. It is related to the dimensionality in terms of the hydrogenic Bohr radius, explained with the equivalence of the proposed FCP model to an infinite QW model and, in fact is not directly related to the physical dimensions. It has been shown that $\beta$ = 3 and 2.55 $\leq \beta \leq$ 2.7 represent the bulk and 2D screening, respectively. The Coulomb potential in fractional space provides an additional parameter in the form of $\beta$ in addition to $\mu$ and $\epsilon$, which allows to incorporate screening correction in material systems induced due to varying thickness, substrates, stress layers and practical conditions under which experimental and numerical calculations are performed. 
 
\begin{figure}[ht!]
\begin{centering}
    \includegraphics[width=1.0\columnwidth]{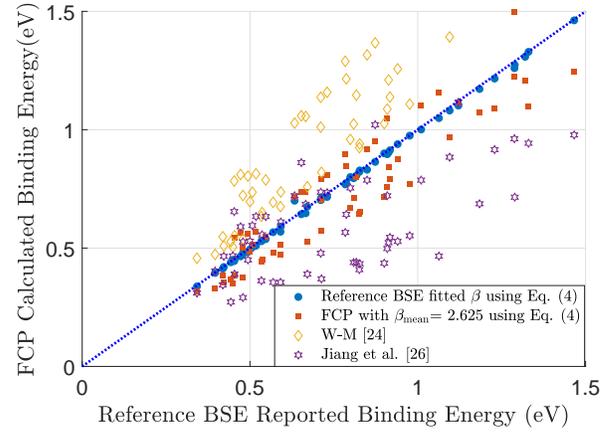}
    \caption{Comparative analysis of relative error MSE in calculated binding energies for FCP model with $\beta_{\text{mean}}$ = 2.625 using $(\ref{equation5})$, W-M and Jiang et al. \cite{PhysRevLett.118.266401} with respect to reported BSE data fitted with respective $\beta$ using $(\ref{equation5})$, taken as reference. FCP calculated binding energies with $\beta_{\text{mean}}$ = 2.625 capable of capturing most of the data in comparatively good agreement up to 1.5 eV in contrast to 0.5 eV for the other two models.}
    \label{figure5}
\end{centering}
\end{figure}

\begin{figure}[ht!]
\begin{centering}
    \includegraphics[width=1.0\columnwidth]{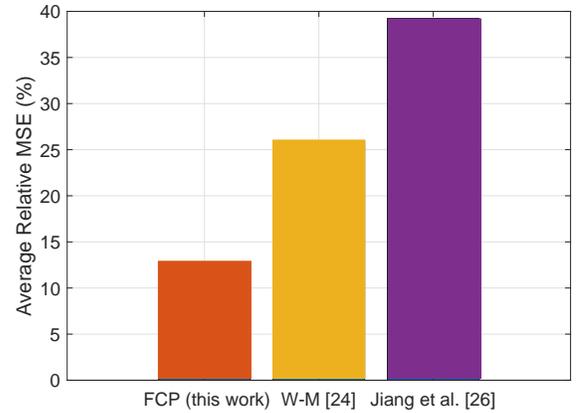}
    \caption{FCP calculation with $\beta_{\text{mean}}$ = 2.625 reduces the relative average MSE for 58 mono-layer materials up to one-half, and one-third in comparison to Jiang et al. \cite{PhysRevLett.118.266401} and  W-M respectively. Moreover a comparison with Olsen et al. \cite{olsen2016simple} shows an approximate reduction in the error up to one-fourth (see the Supplemental Material for details \cite{SUPPLIMENTAL}).}
    \label{figure6}
\end{centering}
\end{figure}

 The result in Fig. \ref{figure5} illustrates an error analysis of the proposed FCP model in comparison with existing models by using BSE reported data for 58 mono-layer materials as reference. The BSE data is calculated accurately by the FCP model fitted with respective $\beta$, whereas, the W-M data having an inherently fixed $\beta$ = 2.515, shows an overestimation increasing with the binding energies. However, the exciton model by Jiang et al. \cite{PhysRevLett.118.266401} has a variable $\beta$ and is shown to exhibit an increase in underestimation for binding energies greater than 0.5 eV. Moreover, the existing models capture reported values in good agreement limited up to 0.5 eV, also reported by Olsen et al. \cite{olsen2016simple}. This limitation is caused due to the incorrect dielectric screening in existing models, which is validated by the FCP model with the identification of actual average screening represented at $\beta_{\text{mean}}$ = 2.625. This corrected $\beta$ is found to result in binding energies, to comparatively fall in good agreement with the reported BSE data up to 1.5 eV. The result in Fig. \ref{figure6} shows an average reduction in the relative mean square error (MSE) for the FCP model with $\beta_{\text{mean}}$ = 2.625 to less than 12.8$\%$, which is approximately one-third and one-half in comparison to W-M and Jiang et al. \cite{PhysRevLett.118.266401}, respectively.

 It is to be noted that the simple screened hydrogen model with classical Coulomb potential of $1/r$ dependence demonstrates W-M excitons of delocalized nature and hence can model binding energies in few hundreds of meV correctly. Since the dielectric screening in some 2D materials can be greatly reduced, the exciton binding energy may reach higher values (up to 1 eV and above). However, their corresponding quasi-classical radius of the ground-state orbital may still remain substantially larger than the lattice constant~\cite{trushin2019tightly}; such excitons can be regarded as W-M excitons and sometimes referred to as tightly bound 2D W-M excitons. Here the materials with higher binding energies are presumed to fall into this category of tightly bound 2D W-M excitons, and our FCP model of $1/r^{\beta-2}$ dependence can model them correctly. Once the exciton radius becomes comparable to the lattice constant, the very W-M picture becomes inapplicable and a non-Coulombic electron-hole interaction takes place~\cite{trushin2019tightly,pelant2012luminescence}, which is beyond the scope of the current study. Moreover, binding energies smaller than 0.35 eV are not experimented due to limited availability of the reference data; the materials included in this work have high dynamic and thermodynamic stability with available reference values of BSE binding energy, $G_{0}W_{0}$ bandgap, effective mass and 2D polarizability~\cite{haastrup2018computational}.

 The FCP model intercepts BSE reported 2D binding energies based on a hydrogenic solution with the Coulomb potential represented in the fractional space by including an effective linear screening based on 2D polarizability and additionally allows for the correction in 2D screening through $\beta$, resulting in a Bohr radius corresponding to a 2D regime. Further, an exciton Bohr radius of $\SI{5.53}{\angstrom}$ is calculated for mono-layer MoS$_{2}$ as R$_{exc}$ = $a_{\beta} \times (\epsilon/\mu$), corresponding to a binding energy of 0.547 eV, intercepted at $\beta$ = 2.631. Here $\mu = 0.236$, $\epsilon = 4.003$ and $a_{\beta} = \SI{0.326}{\angstrom}$ is calculated as $w/19.75$, where $w$ is equal to $\SI{6.44}{\angstrom}$ computed at $\beta = 2.631$ using the result in Fig. \ref{figure3}. The calculated value of $\SI{5.53}{\angstrom}$ is in good agreement with theoretically reported value of $\approx$ \SI{5.5}{\angstrom} \cite{jia2016excitonic} corresponding to an experimentally reported mono-layer thickness of $\sim$0.65 nm \cite{li2015two}. Although an exciton radius of 1 nm is also reported~\cite{qiu2013optical,berkelbach2013theory} corresponding to varying values of binding energies for freestanding mono-layers as 0.96 eV and 0.54 eV, they do not explain why the exciton radius does not change. Because a change in the radius is linked to the screening \cite{li2018correction}, and the screening effect has been shown to change the binding energies \cite{berghauser2014analytical}. We believe that such discrepancies can arise due to different approaches opted to initiate numerical calculations, and a further investigation in the future will prove to be fruitful.

\begin{figure}[ht!]
\begin{centering}
    \includegraphics[width=1.0\columnwidth]{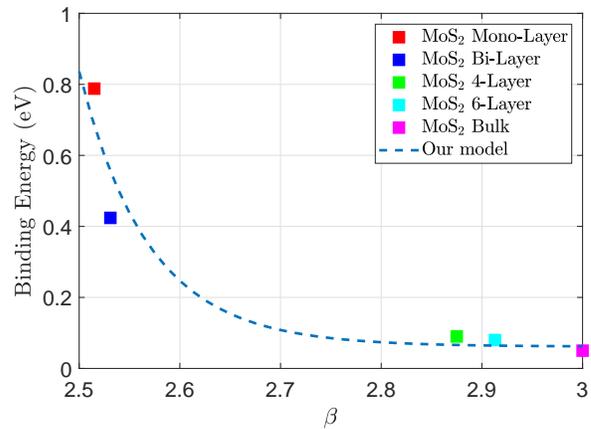}
    \caption{\textcolor{black}{Demonstration of smooth transition of $\beta$ with respect to binding energy} for MoS$_{2}$, given $\mu$ and $\epsilon$ data for the respective number of layers. Blue dotted line represents a power law fit to the data points given in Table \ref{tab:table main1}.}
    \label{figure7}
\end{centering}
\end{figure}

\begin{figure}[ht!]
\begin{centering}
    \includegraphics[width=1.0\columnwidth]{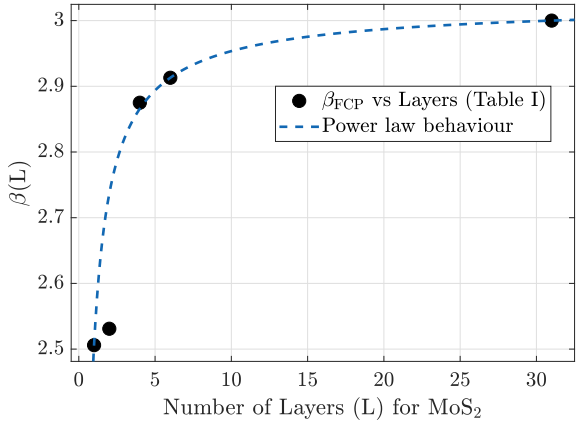}
    \caption{\textcolor{black}{Demonstration of the smooth transition of $\beta$ with respect to the number of layers in MoS$_{2}$, showing the usefulness of the FCP model in the design process for practical applications. \textcolor{black}{Blue dotted line shows $\beta(\text{L})$ to follow a power law model as $(c_{1}\text{L}^{c_{2}}+c_{3})$ which is obtained by performing a fit to the data points given in Table \ref{tab:table main1}, where $c_{1} = -0.5233$, $c_{2} = -0.8394$ and $c_{3} = 3.029$ with confidence bounds of (-1.19, 0.1439), (-3.385, 1.707) and (2.413, 3.646), respectively. Black markers indicate the FCP data.}}} 
    \label{figure8}
\end{centering}
\end{figure}

  A smooth transition of $\beta$ with structural confinement from bulk to mono-layer, given the material data for the respective number of layers, is presented in Fig. \ref{figure7}. \textcolor{black}{This demonstration is based on a power law fit to the reported MoS$_{2}$ binding energies corresponding to a decreasing number of layers, intercepted with the FCP model through $\beta$ as given in Table \ref{tab:table main1}}. \textcolor{black}{In practice, the variation of binding energy is inversely proportional to the dielectric constant, whereas dielectric constant is known to have a direct relationship with the structural confinement \cite{li2016study,liu2019energy}. Due to which the proposed FCP model can provide an approximate estimate of the layer thickness corresponding to the desired binding energy through the mapping of the fractional-dimension parameter $\beta$ as demonstrated in Fig. \ref{figure8}. The result in Fig. \ref{figure8} is an extension of Fig. \ref{figure7}, which shows a relationship between fractional-dimension parameter $\beta$ and the number of layers (L), given by $\beta$(L). \textcolor{black}{Here $\beta(\text{L})$ follows a power law model of the form $(c_{1}\text{L}^{c_{2}}+c_{3})$ and indicates a sharp decrease in the screening for the 2D regime (very small L) compared to the bulk regime (large L), attributed to the strong excitonic effects due to structural confinement. This allows an approximate calculation of the binding energies for changing number of layers as $E_{\beta,\text{L}} = E_{0}\times \big (\,a\beta(\text{L})^{b}+c\big )\,\big (\,\frac{\mu}{\epsilon^{2}}\big)$ where $a$, $b$ and $c$ remain the same as in $(\ref{equation5})$}, highlighting the usefulness of the FCP model in practical design problems. } It is worthwhile to mention that an earlier model reported by Thilagam has reported the kinetic energy part of two-particle fractional-dimensional Schrodinger equation with classical Coulomb potential \cite{thilagam2014exciton}. Contrarily, our FCP model demonstrates that the fractional Coulomb potential corresponds to structural changes from bulk to the 2D regime and thus provides a better representation of dielectric screening in confined materials. 
  
\begin{table}[ht!]
\caption{\label{tab:table main1}
\textcolor{black}{Calculation details for FCP model based interception of reported MoS$_{2}$ binding energies corresponding to a structural confinement from bulk to the mono-layer.}}
\begin{ruledtabular}
\begin{tabular}{cccccc}

Structure&$\mu$&$\epsilon$&Binding Energy &Layers (L)&$\beta_{\text{FCP}}\footnotemark[1]$\\
\hline
Bulk&0.4 \cite{fortin1975excitons}&10.71 \cite{fortin1975excitons}&0.04 eV \cite{fortin1975excitons}&31\footnotemark[2]&3\\

6-Layer&0.3 \cite{kumar2012tunable}&7.92 \cite{kumar2012tunable}&0.08 eV \cite{kumar2012tunable}&6 \cite{kumar2012tunable}&2.913\\

4-Layer&0.25 \cite{kumar2012tunable}&7.16 \cite{kumar2012tunable}&0.09 eV \cite{kumar2012tunable}&4 \cite{kumar2012tunable}&2.875\\

Bi-Layer&0.25 \cite{kumar2012tunable}&5.51 \cite{kumar2012tunable}&0.424 eV \cite{kumar2012tunable}&2\cite{kumar2012tunable}&2.531\\

Mono-Layer&0.19 \cite{cheiwchanchamnangij2012quasiparticle}&3.43 \cite{cheiwchanchamnangij2012quasiparticle}&0.897 eV \cite{cheiwchanchamnangij2012quasiparticle}&1 \cite{cheiwchanchamnangij2012quasiparticle}&2.506\\
\end{tabular}
\end{ruledtabular}
\footnotemark[1]{FCP model interception of binding energies for the respective layers through tuning of $\beta$.}
\footnotemark[2]{$\frac{\text{Bulk Thickness ($\approx$ 20 nm)}}{\text{Mono-Layer Thickness (0.65 nm)}}$ \cite{li2015two}.} 
\end{table}

  \textcolor{black}{As an example for many applications, recent studies on a number of photovoltaic systems based on 2D materials \cite{zhao2017design,massicotte2018dissociation,pospischil2014solar,godefroid2017design} are demonstrated to highlight the influence of various dielectric environments and structural configurations on the critical role of binding energy in solar applications for a favourable dissociation of photo-generated excitons into free carriers at room temperature. Further, 2D binding energy is reported to have a significant contribution to the band gap renormalization and is shown to vary due to the screening induced by the external dielectric environment leading to shifts in the optical spectrum \cite{zhang2020tungsten,qiu2019giant,raja2017coulomb}. Thus 2D material based optical detection which requires a strict optical frequency and narrow line width would demand an accurate determination of binding energy. Here the FCP model will prove to be a promising tool due to its ability to produce accurate binding energies for a wide range of 2D materials with the average screening felt by 2D excitons corrected to represent a more realistic dielectric environment, hence contributing towards the accurate determination of optoelectronic properties. Additionally, it has the potential to track the variation in binding energy due to a change in material thickness from mono-layer to many layers and thus can allow for binding energy tunability to achieve efficient photo-conversion as reported in literature \cite{guo2020two,zhang2018determination}.}   
  
  Furthermore, strain in 2D materials is shown to induce band gap shifts with excitonic effects~\cite{shi2013quasiparticle,moon2019dynamic,waldecker2019rigid}, and various experimental and numerical studies~\cite{ aslan2018probing, feierabend2017impact, song2017quasiparticle} have been reported recently to account for strained excitonic effects. In future, our proposed FCP model model can be extended to calculate the strained binding energies by establishing a physical relationship of $\beta$ with the lattice parameters corresponding to the strain levels. Such an extension can potentially be used as a tool to control optical properties of strained 2D materials~\cite{mueller2018exciton}.

\section{Conclusion}
In conclusion, the proposed model gives a generalized analytical expression to calculate binding energies for a wide range of materials ranging from bulk to 2D regime using the reported material parameters $\mu$ and $\epsilon$. The fractional-dimension parameter $\beta$ = 3 and 2.55 $\leq \beta \leq$ 2.7 corresponds to the actual dielectric screening in bulk and 2D materials, respectively and thus results in an accurate calculation of binding energies for 58 mono-layer and 8 bulk materials benchmarked with reported data. An average screening in 2D excitons represented by $\beta_{\text{mean}}$ = 2.625, comparatively reduces the relative mean square error on average up to one-third with most of the materials captured by our model up to 1.5 eV in contrast to 0.5 eV in existing models. \textcolor{black}{Finally, for a given material, we show that there exists a scaling law between $\beta$ and structural confinement from bulk to mono-layer, which assists in the tuning of binding energy by varying the material thickness. The proposed FCP model will prove useful in the design and engineering of optoelectronic devices based on 2D heterostructures, owning to the critical role of binding energy in the efficient photo-conversion operation. \textcolor{black}{The FCP model can provide a useful tool with its analytical approach through providing accurate binding energy estimates for 2D materials under the impact of practical dielectric environments, thus contributing towards the optimal design of optoelectronic devices.}}

\section{Acknowledgments}
SA and MZ contributed equally to this work. SA is supported by the Information Technology University (ITU) doctoral fellowship. MZ is supported by the ITU start-up grant. MZ acknowledges the travel support for summer visit under the Singapore MOE T2 grant (2018-T2-1-007). LKA would like to acknowledge the support of ONRG grant (N62909-19-1-2047). MZ is thankful to Yee Sin Ang for enlightening conversations.

\setcounter{equation}{0}
\setcounter{figure}{0}
\renewcommand{\theequation}{A\arabic{equation}}
\renewcommand{\thefigure}{A\arabic{figure}}
\appendix*
\section{Detailed derivation and full solution of screened hydrogen model with fractional Coulomb potential (FCP)}

The FCP model is developed based on an arbitrary potential embedded in the infinite quantum well, as illustrated in Fig. \ref{figureA1}, following the eigenvalue calculation method reported in \cite{jugdutt2013solving}.

The radial part of the simple hydrogen model in spherical coordinates incorporating the fractional Coulomb potential represents the FCP model as:
\begin{figure}[ht!]
\begin{centering}
    \includegraphics[width=1.0\columnwidth]{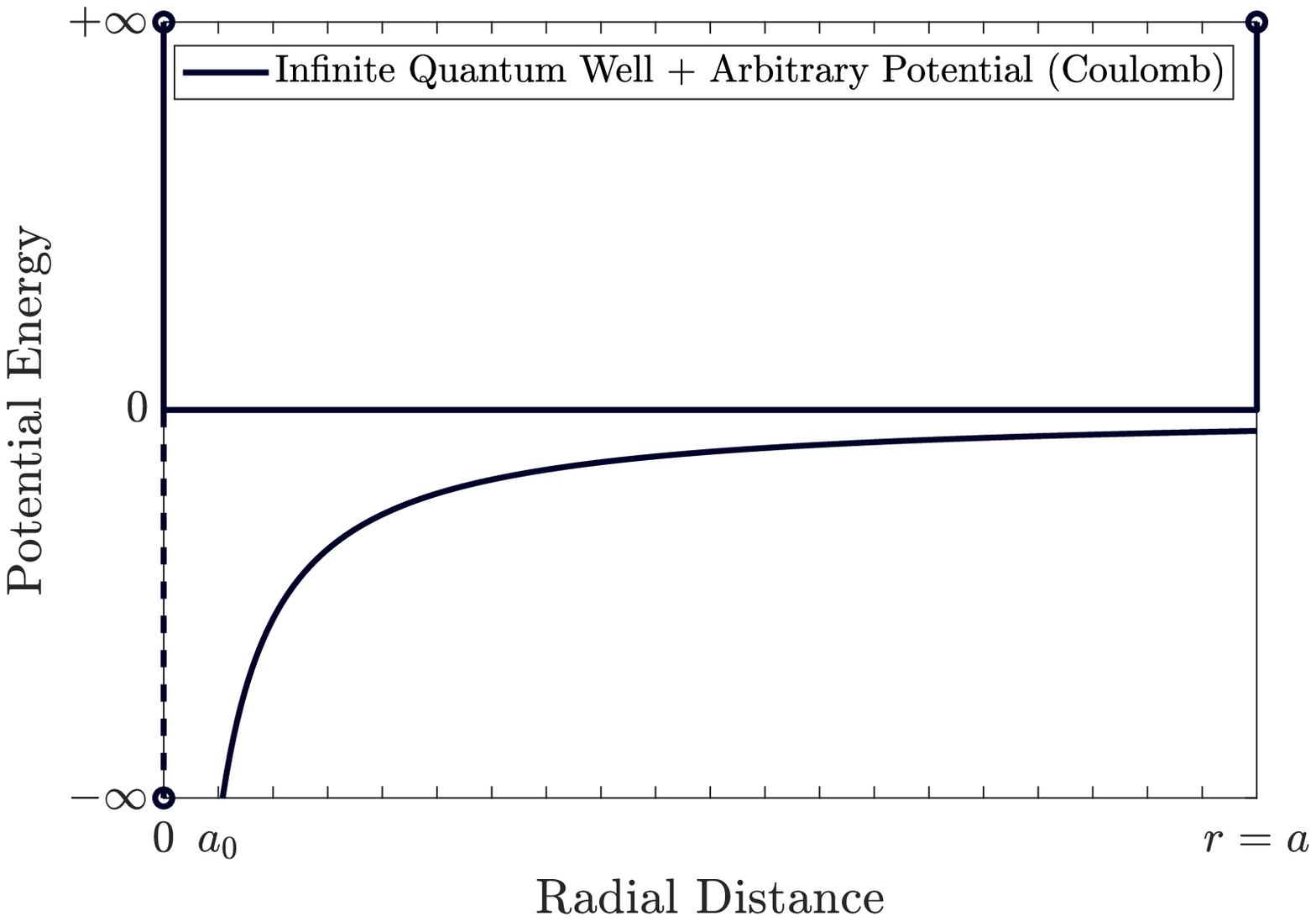}
    \caption{Model framework for hydrogenic energy computation with Coulomb potential embedded in an infinite quantum well. Here $a_{0}$ is the Bohr radius and $a$ is the radial distance of the infinite QW taken at least 20 times of $a_{0}$ to achieve convergence. Here we replace the Coulomb potential with a fractional Coulomb potential.}
    \label{figureA1}
\end{centering}
\end{figure}

\begin{equation}
\begin{aligned}
\bigg ( \, -\frac{\hbar^{2}}{2\mu}\frac{d^{2}}{dr^{2}} + V_\text{frac}(r) \bigg )\,\psi(r) = E\psi(r)
\end{aligned}
\label{equationA1}
\end{equation}
where $\mu$ = 0.9995$m_{0}$ is the exciton reduced mass, $V_\text{frac}(r) = k_{\beta}e^{2}/r^{\beta-2}$ is the fractional Coulomb potential, $r$ is the radial distance equivalent to the infinite QW width and $\beta$ is the fractional-dimension parameter linked to the screened Coulomb potential.

The formulation of the eigenvalue problem for a fractional Coulomb potential embedded in an infinite QW is given as: $\sum_{m=1}^{n_\text{max}}H_{nm}c_{m} = Ec_{n}$ and the Hamiltonian matrix elements are given by:

\begin{equation}
\begin{aligned}
H_{nm} = \langle\phi_{n}|\big(\,H_{0} + V_\text{frac}(r)|\phi_{m}\rangle\big)\,  = \\
\delta_{nm}E_{n}^{0} + V_\text{frac}(r)   
\end{aligned}
\label{equationA2}
\end{equation}
where, $H_{0}|\phi_{n}\rangle = E_{n}^{0}|\phi_{n}\rangle$ represents the infinite QW. The Eigen states of the infinite QW, $\phi_{n}(r) = \sqrt[]{\frac{2}{a}}\sin(n \pi r/a)$ with eigenvalues $E_{n}^{0} = \pi^{2}\hbar^{2}n^{2}/2m_{0}a^{2}$ allow for the Fourier series expansion of the embedded wave function as $|\psi\rangle = \sum_{m=1}^{n_\text{max}}c_{m}|\phi_{m}$ where $\int_{0}^{\infty}|\psi(r)|^{2} dr = 1$. The infinite QW enforces the embedding wave function $\psi(r)$ to satisfy the Von-Karman boundary conditions as $\psi(0) = 0$ and $\psi(r = a) = 0$. Here $m_{0} = 9.11 \times 10^{-31}$ kg and $a = 20\times a_{0}$ is the width of the infinite quantum well. The fractional Coulomb potential over the radial distance is expressed as:

\begin{equation}
\begin{aligned}
V_\text{frac}(r) =  -\frac{2k_{\beta}e^{2}}{a}\int_{0}^{a}\sin\bigg(\,\frac{n\pi r}{a}\bigg)\,\frac{1}{r^{\beta-2}}\sin\bigg(\,\frac{m\pi r}{a}\bigg)\, dr
\end{aligned}
\label{equationA3}
\end{equation}
where $k_{\beta} = \Gamma(\beta/2)/(2\pi^{\beta/2}(\beta-2)\epsilon_{0})$. After trigonometric transformations we arrive at:

\begin{equation}
\begin{aligned}
V_\text{frac}(r) = -\frac{k_{\beta}e^{2}}{a}\bigg(\,G(n+m)-G(n-m) \bigg )\,
\end{aligned}
\label{equationA4}
\end{equation}
where
\begin{equation}
\begin{aligned}
G(n+m) = \int_{0}^{a}\bigg [\,\frac{1 - \cos \big(\,\frac{\pi r}{a}(n+m)\big)\,}{r^{\beta-2}}\bigg]\,dr\\
G(n-m) = \int_{0}^{a}\bigg [\,\frac{1 - \cos \big(\,\frac{\pi r}{a}(n-m)\big)\,}{r^{\beta-2}}\bigg]\,dr
\end{aligned}
\label{equationA5}
\end{equation}

\begin{figure}[ht!]
\begin{centering}
    \includegraphics[width=1.0\columnwidth]{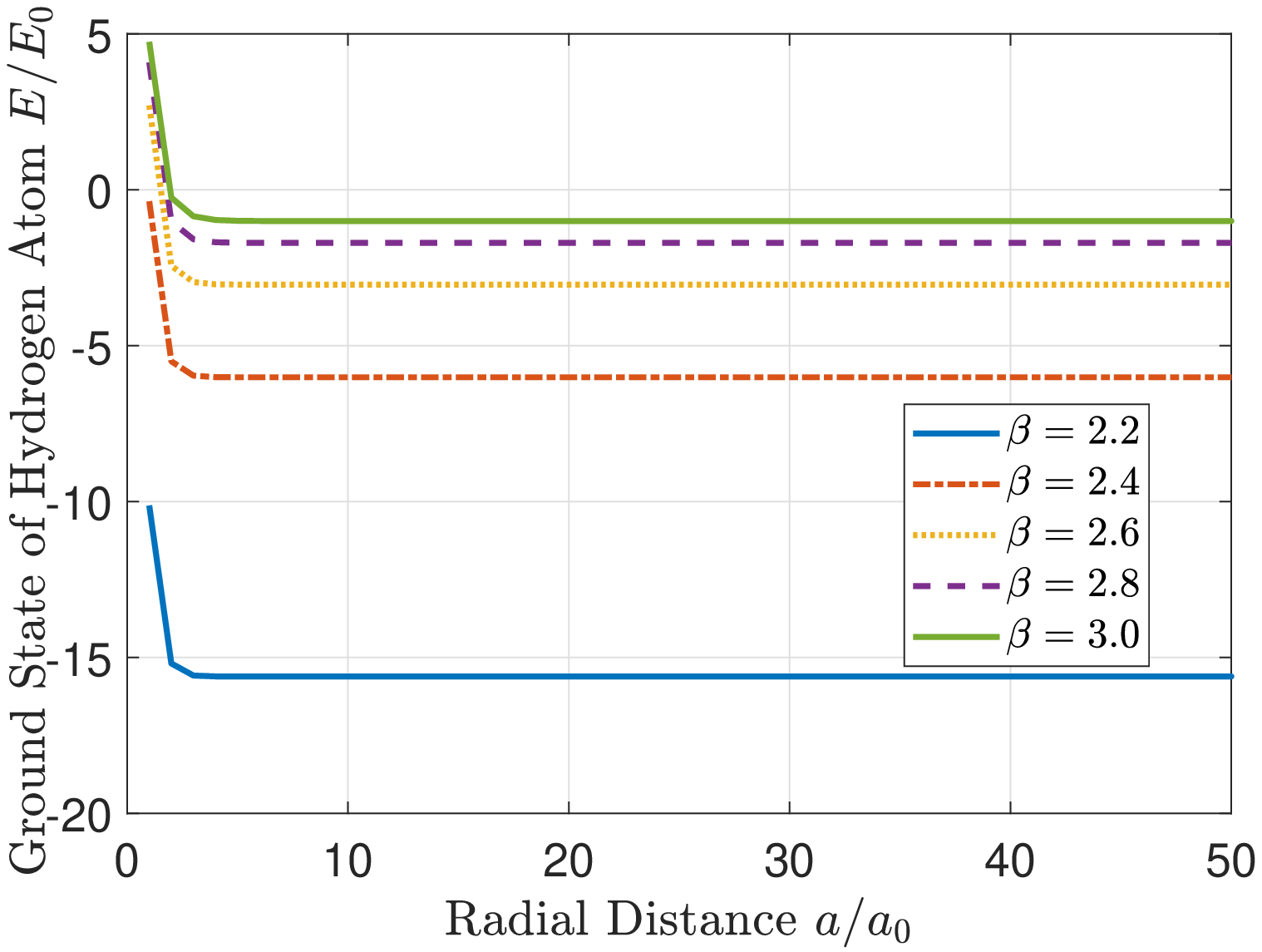}
    \caption{FCP model convergence calculated by varying the radial distance $a/a_{0}$ for $n_{\text{max}} = 400$, $E_{0} = 13.6$ eV and $a_{0} = \SI{0.529}{\angstrom}$.}
    \label{figureA2}
\end{centering}
\end{figure}

\begin{figure}[ht!]
\begin{centering}
    \includegraphics[width=1.0\columnwidth]{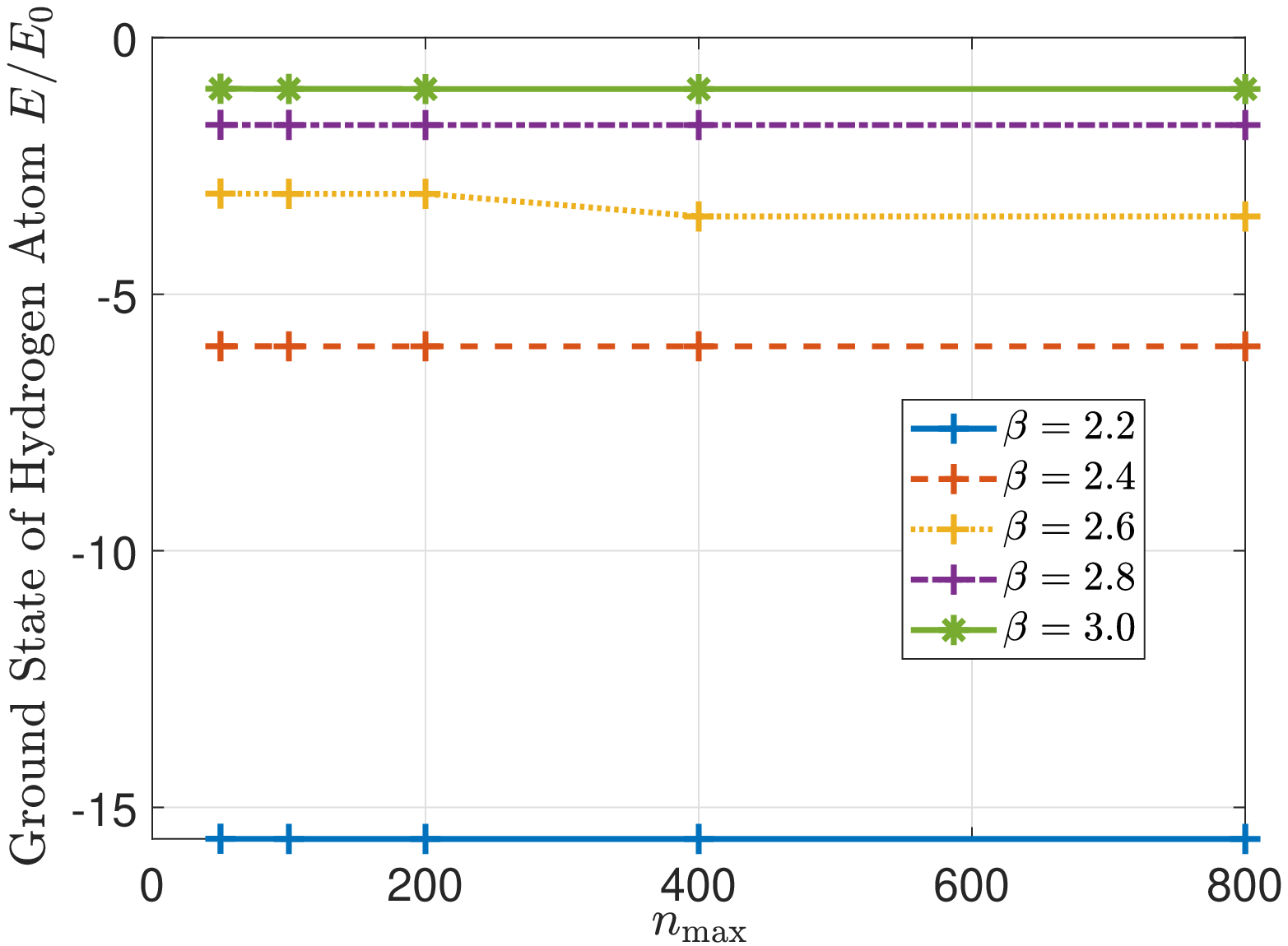}
    \caption{FCP model convergence calculated by varying $n_{\text{max}}$ for $a/a_{0}$ = 20.}
    \label{figureA3}
\end{centering}
\end{figure}

The dimensionless version of the Hamiltonian matrix equation for the FCP model is represented as:
\begin{equation}
\begin{aligned}
\sum_{m=1}^{n_\text{max}}h_{nm}c_{m} = ec_{n}
\end{aligned}
\label{equationA6}
\end{equation}
where $h_{nm} = \frac{H_{nm}}{E_{0}}$, $e = \frac{E}{E_{0}}$ and $E_{0} = \frac{\hbar^{2}}{2m_{0}a_{0}^{2}} \approx 13.606$ eV resulting in an expression for the numerical calculation of eigenvalues as given below:

\begin{equation}
\begin{aligned}
h_{nm} = \delta\bigg(\frac{\pi n a_{0}}{a}\bigg)^{2}\\
-\frac{4\pi ^{(1-\frac{\beta}{2})}\Gamma(\beta/2)}{(\beta-2)}\frac{a_{0}}{a}\bigg[G(n+m)-G(n-m)\bigg]
\end{aligned}
\label{equationA7}
\end{equation}

The integrals are computed using trapezoidal integration and the dimensionless Hamiltonian matrix for FCP model is solved for the first bound states by employing block diagonalization using Jacobian rule.  

The plots in Fig. \ref{figureA2} and Fig. \ref{figureA3} show the convergence of the FCP model for the ground state energy for $2 < \beta \leq 3$ with respect to $a/a_\text{0}$ and $n_\text{max}$ respectively. Where $n_\text{max}$ represents the Hamiltonian matrix size dependent on the cutoff energy of the wave function and $a/a_\text{0}$ represents the radial distance equivalent to an infinite QW width which is important for convergence. We have performed our calculation for $n_\text{max}$ = 400 and $a/a_\text{0}$ = 50.

\setcounter{equation}{0}
\setcounter{figure}{0}
\renewcommand{\theequation}{B\arabic{equation}}
\renewcommand{\thefigure}{B\arabic{figure}}
\bibliography{Main}
\pagebreak
\onecolumngrid
\begin{center}
	\textbf{\large Supplementary Material\\Generalized scaling law for exciton binding energy in two-dimensional materials}\\
\end{center}	

\setcounter{equation}{0}
\setcounter{figure}{0}
\setcounter{table}{0}
\renewcommand{\theequation}{S\arabic{equation}}
\renewcommand{\thefigure}{S\arabic{figure}}
\renewcommand{\thefigure}{S\arabic{figure}}
\renewcommand{\thetable}{S\arabic{table}}

\begin{figure}[h]
	\begin{centering}
		\includegraphics[width=0.8\columnwidth]{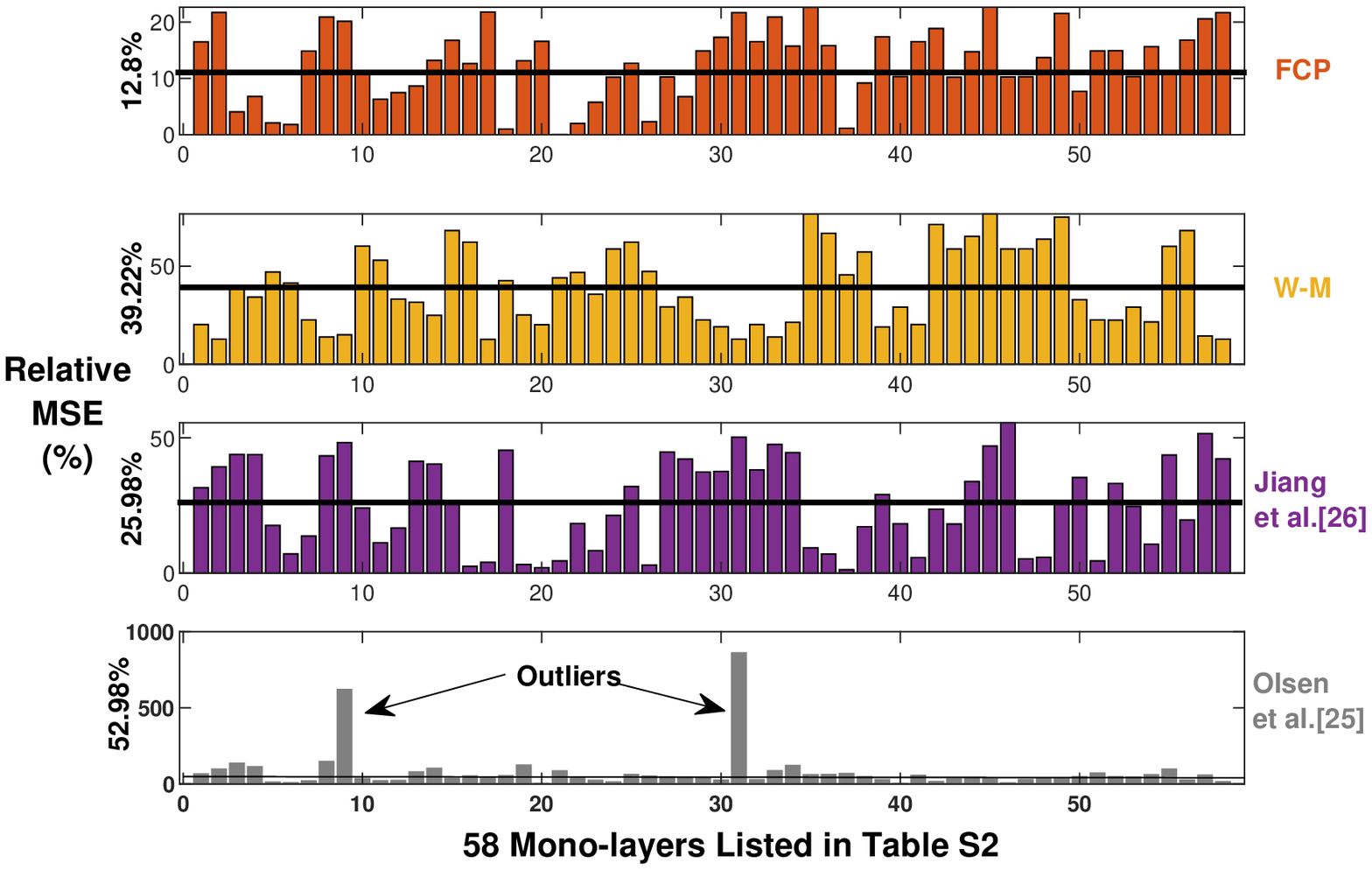}
		\caption{Relative mean square error (MSE) analysis for the proposed FCP model at $\beta_{\text{mean}} = 2.625$ in comparison to other models with calculations performed on 58 mono-layer materials, the details of which are given in Table \ref{tab:table2}. The numbers on the Y-axis show the average relative MSE of 12.8 $\%$, 39.22 $\%$, 25.98 $\%$ and 52.98 $\%$ corresponding to FCP model calculations at $\beta_{\text{mean}}$ = 2.625, FCP calculations for Wannier-Mott \cite{wannier1937structure,cheiwchanchamnangij2012quasiparticle}, FCP calculations for Jiang et al \cite{PhysRevLett.118.266401} and FCP calculations for Olsen et al. \cite{olsen2016simple}, respectively. Note that in case of Olsen et al. \cite{olsen2016simple}, the average relative MSE is calculated by excluding the outliers listed at the 9$^{th}$ and 31$^{st}$ place in Table \ref{tab:table2}.}
		\label{figuresuup1}
	\end{centering}
\end{figure}

\begin{table}[b]
	\caption{\label{tab:table1}
		FCP model based calculation of BSE reported binding energies for 58 mono-layer materials \cite{haastrup2018computational} through $\beta$.}
	\begin{ruledtabular}
		\footnotesize
		\resizebox{0.8\columnwidth}{!}{
			\begin{tabular}{ccccccccc}
				&Monolayer&$\mu_{x}/\mu_{y}/\mu\footnotemark[1]$&$\alpha_{x}/\alpha_{y}/\alpha\footnotemark[2]$&$\epsilon_{\text{eff}}\footnotemark[3]$&E$_{\text{BSE}}\footnotemark[4]$(eV)&E$_{\text{FCP}}\footnotemark[5]$(eV)&$\beta_{\text{FCP}}\footnotemark[6]$&E$_{\text{A}}\footnotemark[7]$(eV)\\  
				\hline
				1 & CrS$_{2}$-H&0.439/0.438/0.438&8.562/8.562/8.562&6.1305&0.534&0.532&2.569&0.529\\
				
				2&CrSe$_{2}$-H&0.473/0.472/0.472&10.309/10.309/10.309&6.9075&0.479&0.479&2.55&0.479\\
				
				3&HfS$_{2}$-H&2.290/0.705/1.497&3.155/3.155/3.155&6.8111&1.289&1.285&2.612&1.274\\
				
				4&HfSe$_{2}$-H&1.793/0.209/1.001&4.380/4.380/4.380&6.5811&0.94&0.94&2.603&0.939\\
				
				5&MoS$_{2}$-H&0.236/0.236/0.236&6.192/6.192/6.192&4.0344&0.547&0.545&2.631&0.538\\
				
				6&MoSe$_{2}$-H&0.267/0.267/0.267&7.28/7.28/7.28&4.566&0.5&0.5&2.619&0.494\\
				
				7&MoTe$_{2}$-H&0.27/0.27/0.27&9.468/9.468/9.468&5.1546&0.454&0.454&2.575&0.452\\
				
				8&TiS$_{2}$-H&1.762/0.653/1.207&5.099/5.099/5.099&7.6993&0.976&0.976&2.553&0.975\\
				
				9&TiSe$_{2}$-H&4.582/0.569/2.575&7.311/7.311/7.311&13.069&0.713&0.715&2.556&0.715\\
				
				10&WS$_{2}$-H&0.166/0.166/0.166&5.577/5.577/5.577&3.3294&0.518&0.5177&2.659&0.51\\
				
				11&WSe$_{2}$-H&0.186/0.186/0.186&6.590/6.590/6.590&3.7432&0.479&0.48&2.644&0.473\\
				
				12&WTe$_{2}$-H&0.160/0.160/0.160&8.859/8.859/8.859&3.982&0.419&0.4174&2.6&0.413\\
				
				13&ZrS$_{2}$-H&1.618/0.483/1.050&3.527/3.527/3.527&6.0937&1.184&1.182&2.597&1.172\\
				
				14&ZrSe$_{2}$-H&1.373/0.474/0.923&4.844/4.844/4.844&6.6421&0.917&0.918&2.581&0.913\\
				
				15&GeS$_{2}$-T&0.935/0.117/0.526&3.966/3.966/3.966&4.7102&0.784&0.7827&2.675&0.769\\
				
				16&HfS$_{2}$-T&0.168/0.166/0.167&3.604/3.604/3.604&2.8004&0.73&0.7293&2.663&0.716\\
				
				17&NiS$_{2}$-T&0.483/0.128/0.305&10.609/10.609/10.609&5.7347&0.449&0.449&2.55&0.449\\
				
				18&PbS$_{2}$-T&1.066/0.179/0.622&4.619/4.619/4.619&5.4333&0.817&0.8158&2.622&0.806\\
				
				19&PdS$_{2}$-T&1.164/0.160/0.662&7.544/7.544/7.544&6.9875&0.592&0.594&2.581&0.591\\
				
				20&PdSe$_{2}$-T&0.268/0.116/0.192&10.799/10.799/10.799&4.6976&0.396&0.396&2.569&0.395\\
				
				21&PtS$_{2}$-T&1.058/0.255/0.656&5.335/5.335/5.335&5.9398&0.711&0.713&2.625&0.704\\
				
				22&PtSe$_{2}$-T&0.524/0.166/0.345&7.068/7.068/7.068&5.0473&0.507&0.509&2.631&0.503\\
				
				23&PtTe$_{2}$-T&0.159/0.100/0.129&10.575/10.575/10.575&3.9238&0.342&0.3413&2.606&0.338\\
				
				24&SnS$_{2}$-T&0.976/0.156/0.566&3.168/3.168/3.168&4.4079&1.01&1.01&2.656&1.001\\
				
				25&SnSe$_{2}$-T&1.008/0.134/0.571&4.615/4.615/4.615&5.225&0.711&0.716&2.663&0.703\\
				
				26&ZrS$_{2}$-T&0.176/0.164/0.17&4.207/4.207/4.207&2.9983&0.633&0.63&2.632&0.7\\
				
				27&HfBr$_{2}$&0.794/0.369/0.581&5.195/5.195/5.195&5.5554&0.809&0.8025&2.591&0.795\\
				
				28&HfCl$_{2}$&0.792/0.498/0.645&4.367/4.367/4.367&5.3833&0.913&0.9119&2.60&0.904\\
				
				29&HfI$_{2}$&0.168/0.168/0.168&6.785/6.785/6.785&3.6304&0.569&0.57&2.575&0.567\\
				
				30&TiBr$_{2}$&0.357/0.357/0.357&7.814/7.814/7.814&5.36&0.591&0.5899&2.566&0.569\\
				
				31&TiCl$_{2}$&4.824/3.077/3.950&6.560/6.560/6.560&15.243&0.826&0.8235&2.55&0.822\\
				
				32&TiI$_{2}$&0.282/0.281/0.281&10.021/10.021/10.021&5.3869&0.443&0.4425&2.569&0.44\\
				
				33&ZrBr$_{2}$&0.842/0.708/0.775&5.871/5.871/5.871&6.6942&0.827&0.8292&2.553&0.828\\
				
				34&ZrCl$_{2}$&1.014/0.976/0.995&5.127/5.127/5.127&7.0564&0.909&0.9026&2.572&0.898\\
				
				35&BN&0.380/0.380/0.380&0.960/0.960/0.960&2.446&2.001&2.078&2.678&1.962\\
				
				36&BP&0.092/0.092/0.092&4.832/4.832/4.832&2.4935&0.494&0.493&2.672&0.484\\
				
				37&GaN&0.123/0.123/0.123&1.373/1.373/1.373&2.044&1.121&1.126&2.666&1.102\\
				
				38&Al$_{2}$O$_{2}$&0.205/0.205/0.205&3.589/3.589/3.589&3.0294&0.787&0.7866&2.653&0.775\\
				
				39&Al$_{2}$S$_{2}$&0.486/0.472/0.479&3.104/3.104/3.104&4.0645&1.33&1.336&2.566&1.329\\
				
				40&AL$_{2}$Se$_{2}$&0.680/0.364/0.522&3.631/3.631/3.631&4.516&1.095&1.09&2.591&1.081\\
				
				41&AL$_{2}$Te$_{2}$&0.173/0.173/0.173&4.607/4.607/4.607&3.1319&0.799&0.804&2.569&0.8\\
				
				42&Ga$_{2}$O$_{2}$&0.533/0.532/0.532&2.196/2.196/2.196&3.6696&1.288&1.282&2.681&1.26\\
				
				43&Ga$_{2}$S$_{2}$&0.288/0.288/0.288&3.345/3.345/3.345&3.3845&0.873&0.877&2.656&0.864\\
				
				44&Ga$_{2}$Se$_{2}$&0.195/0.194/0.194&4.189/4.189/4.189&3.16&0.653&0.654&2.669&0.643\\
				
				45&Ga$_{2}$Te$_{2}$&0.132/0.132/0.132&5.548/5.548/5.548&3.0268&0.453&0.454&2.691&0.445\\
				
				46&In$_{2}$O$_{2}$&0.487/0.487/0.487&2.930/2.930/2.930&3.9934&1.063&1.066&2.656&1.049\\
				
				47&In$_{2}$S$_{2}$&0.294/0.294/0.294&3.518/3.518/3.518&3.4857&0.849&0.844&2.656&0.831\\
				
				48&In$_{2}$Se$_{2}$&0.201/0.201/0.201&4.235/4.235/4.235&3.2168&0.666&0.659&2.666&0.647\\
				
				49&In$_{2}$Te$_{2}$&0.137/0.136/0.136&5.356/5.356/5.356&3.0248&0.472&0.474&2.688&0.465\\
				
				50&GeS&0.359/0.322/0.340&2.598/2.589/2.598&3.2678&1.32&1.319&2.6&1.308\\
				
				51&Pb$_{2}$S$_{2}$&0.088/0.063/0.075&6.010/6.009/6.0095&2.5127&0.536&0.5348&2.575&0.532\\
				
				52&SnS&0.380/0.377/0.378&2.564/2.564/2.564&3.3948&1.466&1.469&2.575&1.462\\
				
				53&SnSe&0.342/0.322/0.332&2.925/2.925/2.925&3.3957&1.229&1.226&2.591&1.216\\
				
				54&Ge$_{2}$Se${2}$&0.098/0.069/0.083&8.059/8.909/8.484&2.9869&0.422&0.4228&2.572&0.42\\
				
				55&GeO&2.348/0.980/1.682&1.977/1.977/1.977&5.8017&1.731&1.727&2.659&1.702\\
				
				56&S$_{2}$Si$_{2}$&0.985/0.145/0.565&3.356/3.366/3.361&4.5197&0.909&0.913&2.675&0.897\\
				
				57&P$_{4}$&1.357/0.067/0.712&6.055/4.533/5.294&6.1416&0.9&0.9&2.553&0.9\\
				
				58&As$_{4}$&0.671/0.099/0.385&6.357/7.486/6.9215&5.2512&0.671&0.6762&2.55&0.675\\
				
				59&4-layer MoS$_{2}$ \cite{kumar2012tunable}&0.25&---&7.16&0.09&0.917&2.875&---\\
				
				60&6-layer MoS$_{2}$ \cite{kumar2012tunable}&0.3&---&7.92&0.08&0.0822&2.913&---\\
				
				61&Bilayer MoS$_{2}$ \cite{kumar2012tunable}&0.25&---&5.51&0.424&0.4239&2.531&0.4251\\
				
			\end{tabular}
		}
	\end{ruledtabular}
	\footnotetext[1]{$\mu$ is the exciton effective mass, $\mu = \frac{\mu_{x}+\mu_{y}}{2}$, $\mu_{x/y} = \frac{me_{x/y}mh_{x/y}}{me_{x/y}+mh_{x/y}}$ where $me_{x/y}$ and $mh_{x/y}$ are obtained from \cite{haastrup2018computational}.} 
	\footnotetext[2]{$\alpha$ is the static electronic polarizability, $\alpha = \frac{\alpha_{x}+\alpha_{y}}{2}$, where $\alpha_{x/y}$ is obtained from \cite{haastrup2018computational}.}
	\footnotetext[3]{$\epsilon_{\text{eff}}$ is the effective dielectric constant calculated with $\frac{0.5}{(1+\sqrt{1+32\pi\alpha\mu/3})}$ also used by Olsen et al. \cite{olsen2016simple}.}
	\footnotetext[4]{Reported BSE binding energy obtained from \cite{haastrup2018computational}.}
	\footnotetext[5]{Full solution based FCP calculation of reported BSE binding energy through tuning of $\beta$.}
	\footnotetext[6]{FCP model based $\beta$ tuned for intercepting E$_{\text{FCP}}$.}
	\footnotetext[7]{FCP based analytical calculation of binding energy using $\beta_{\text{FCP}}$.}
\end{table}

\begin{table}[b]
	\caption{\label{tab:table2}
		Comparative error analysis of the FCP calculated binding energies at $\beta_{\text{mean}}$ = 2.625 with the FCP calculated binding energies for Wannier-Mott \cite{wannier1937structure,cheiwchanchamnangij2012quasiparticle}, Jiang et al. \cite{PhysRevLett.118.266401} and Olsen et al. \cite{olsen2016simple} using reported values in Table \ref{tab:table1}.}
	\begin{ruledtabular}
		\footnotesize
		\resizebox{0.8\columnwidth}{!}{
			\begin{tabular}{cccccccccccc}
				
				&Monolayers&E$_{\text{g}}\footnotemark[1]$/E$_{\text{U}}\footnotemark[2]$(eV)&$\beta_{\text{U}}\footnotemark[3]$&MSE\footnotemark[4]($\%$)&E$_{\text{WM}}\footnotemark[5]$(eV)&MSE\footnotemark[6]($\%$)&E$_{\text{FCPm}}\footnotemark[7]$(eV)&MSE\footnotemark[8]($\%$)&E$_{\text{OLS}}\footnotemark[9]$(eV)&$\beta_{\text{OLS}}\footnotemark[10]$&MSE\footnotemark[11]($\%$)\\
				
				\hline
				1&CrS$_{2}$-H&1.448/0.362&2.689 &31.569	&0.634	&20.301	&0.441	&16.466	&0.876	&2.425	&65.784\\
				
				2&CrSe$_{2}$-H&1.165/0.291&2.708 &39.248	&0.538	&12.765	&0.375	&21.699	&0.944	&2.362	&97.286\\
				
				3&HfS$_{2}$-H&2.862/0.715&2.810	&43.877	&1.756	&38.201	&1.222	&4.037	&2.994	&2.369	&135.086\\
				
				4&HfSe$_{2}$-H&2.115/0.528&2.798	&43.769	&1.257	&34.252	&0.875	&6.779	&2.001	&2.387	&113.205\\
				
				5&MoS$_{2}$-H&2.533/0.633&2.580	&17.657	&0.788	&47.001	&0.549	&2.072	&0.472	&2.673	&12.267\\
				
				6&MoSe$_{2}$-H&2.119/0.529&2.597	&7.085	&0.696	&41.404	&0.485	&1.813	&0.534	&2.595	&8.097\\
				
				7&MoTe$_{2}$-H&1.563/0.390&2.620	&13.716	&0.552	&22.626	&0.384	&14.852	&0.54	&2.522	&19.469\\
				
				8&TiS$_{2}$-H&2.208/0.552&2.736	&43.384	&1.108	&13.954	&0.771	&20.874	&2.414	&2.306	&147.692\\
				
				9&TiSe$_{2}$-H&1.481/0.370&2.771	&48.251	&0.82	&15.032	&0.571	&20.125	&5.150	&2.060	&620.419\\
				
				10&WS$_{2}$-H&2.532/0.633&2.590	&24.117	&0.814	&60.160	&0.502	&6.292	&0.372	&2.724	&21.353\\
				
				12&WTe$_{2}$-H&1.377/0.344&2.658	&16.707	&0.548	&33.265	&0.382	&7.465	&0.32	&2.682	&22.518\\
				
				13&ZrS$_{2}$-H&2.748/0.687&2.775	&41.382	&1.538	&31.598	&1.070	&8.622	&2.099	&2.428	&79.180\\
				
				14&ZrSe$_{2}$-H&2.183/0.545&2.750	&40.306	&1.138	&24.988	&0.792	&13.212	&1.845	&2.382	&102.190\\
				
				15&GeS$_{2}$-T&2.266/0.566&2.782	&26.397	&1.289	&68.163	&0.897	&16.766	&1.052	&2.576	&36.801\\
				
				16&HfS$_{2}$-T&2.938/0.734&2.655	&2.513	&1.158	&62.223	&0.806	&12.642	&0.334	&2.954	&53.351\\
				
				17&NiS$_{2}$-T&1.868/0.467&2.538	&4.008	&0.504	&12.663	&0.351	&21.770	&0.61	&2.462	&35.857\\
				
				18&PbS$_{2}$-T&1.763/0.440&2.832	&45.409	&1.146	&42.586	&0.797	&0.993	&1.243	&2.492	&54.342\\
				
				19&PdS$_{2}$-T&2.441/0.610&2.571	&3.214	&0.737	&25.134	&0.513	&13.111	&1.323	&2.356	&124.027\\
				
				20&PdSe$_{2}$-T&1.615/0.403&2.563	&2.025	&0.473	&20.144	&0.329	&16.576	&0.384	&2.577	&2.784\\
				
				21&PtS$_{2}$-T&2.945/0.736&2.611	&4.545	&1.011	&44.058	&0.704	&0.028	&1.311	&2.442	&86.363\\
				
				22&PtSe$_{2}$-T&2.381/0.595&2.579	&18.290	&0.736	&46.853	&0.512	&1.969	&0.69	&2.534	&37.176\\
				
				23&PtTe$_{2}$-T&1.242/0.310&2.633	&8.284	&0.457	&35.734	&0.318	&5.750	&0.259	&2.692	&23.372\\
				
				24&SnS$_{2}$-T&3.15/0.787&2.737	&21.378	&1.584	&58.733	&1.103	&10.219	&1.132	&2.616	&13.086\\
				
				25&SnSe$_{2}$-T&1.912/0.478&2.798	&32.005	&1.137	&62.277	&0.792	&12.679	&1.141	&2.514	&62.446\\
				
				26&ZrS$_{2}$-T&2.885/0.721&2.622	&3	&1.028	&47.350	&0.716	&2.314	&0.34	&2.893	&51.428\\
				
				27&HfBr$_{2}$&1.759/0.439&2.791	&44.779	&1.024	&29.271	&0.713	&10.238	&1.162	&2.479	&46.289\\
				
				28&HfCl$_{2}$&2.092/0.523&2.787	&42.146	&1.21	&34.290	&0.842	&6.753	&1.289	&2.497	&42.699\\
				
				29&HfI$_{2}$&1.421/0.355&2.726	&37.389	&0.693	&22.621	&0.482	&14.855	&0.336	&2.746	&40.740\\
				
				30&TiBr$_{2}$&1.421/0.355&2.717	&37.609	&0.675	&19.117	&0.470	&17.288	&0.714	&2.500	&25.483\\
				
				31&TiCl$_{2}$&1.636/0.409&2.779	&50.243	&0.924	&12.806	&0.643	&21.671	&7.899	&1.994	&861.070\\
				
				32&TiI$_{2}$&1.09/0.272&2.724	&38.181	&0.527	&20.253	&0.367	&16.500	&0.563	&2.497	&27.954\\
				
				33&ZrBr$_{2}$&1.737/0.434&2.763	&47.584	&0.94	&13.926	&0.655	&20.893	&1.549	&2.378	&87.198\\
				
				34&ZrCl$_{2}$&1.993/0.498&2.766	&44.543	&1.087	&21.375	&0.756	&15.721	&1.989	&2.351	&121.603\\
				
				35&BN&7.117/1.779&2.724	&9.327	&3.846	&76.571	&2.405	&22.605	&0.76	&3.091	&61.264\\
				
				36&BP&1.802/0.450&2.696	&7.024	&0.804	&66.753	&0.560	&15.787	&0.184	&3.070	&61.983\\
				
				37&GaN&4.465/1.116&2.624	&1.270	&2.087	&45.632	&1.114	&1.121	&0.346	&3.346	&68.548\\
				
				38&Al$_{2}$O$_{2}$&2.568/0.642&2.715	&17.161	&1.215	&57.213	&0.846	&9.163	&0.41	&2.885	&47.096\\
				
				39&Al${2}$S$_{2}$&3.774/0.943&2.674	&29.044	&1.577	&18.999	&1.098	&17.370	&0.958	&2.669	&27.915\\
				
				40&AL$_{2}$Se$_{2}$&3.537/0.884&2.654	&18.223	&1.392	&29.147	&0.969	&10.324	&1.044	&2.601	&3.4227\\
				
				41&AL$_{2}$Te$_{2}$&3.016/0.754&2.587	&5.750	&0.959	&20.251	&0.667	&16.501	&0.346	&2.857	&56.75\\
				
				42&Ga$_{2}$O$_{2}$&3.851/0.962&2.775	&23.650	&2.151	&71.183	&1.497	&18.863	&1.065	&2.738	&15.476\\
				
				43&Ga$_{2}$S$_{2}$&4.084/1.021&2.603	&18.171	&1.367	&58.722	&0.952	&10.211	&0.576	&2.797	&33.333\\
				
				44&Ga$_{2}$Se$_{2}$&3.445/0.861&2.577	&33.903	&1.059	&65.228	&0.737	&14.728	&0.389	&2.850	&39.502\\
				
				45&Ga$_{2}$Te$_{2}$&2.617/0.654&2.569	&46.966	&0.783	&76.602	&0.545	&22.626	&0.264	&2.885	&40.674\\
				
				46&In$_{2}$O$_{2}$&1.867/0.466&2.966	&55.576	&1.661	&58.788	&1.156	&10.256	&0.974	&2.680	&7.149\\
				
				47&In$_{2}$S$_{2}$&3.149/0.787&2.674	&5.294	&1.316	&58.824	&0.916	&10.281	&0.588	&2.775	&29.241\\
				
				48&In$_{2}$Se$_{2}$&2.74/0.685&2.647	&5.873	&1.056	&63.754	&0.735	&13.705	&0.402	&2.836	&37.867\\
				
				49&In$_{2}$Te$_{2}$&2.362/0.590&2.611	&26.881	&0.811	&74.999	&0.565	&21.513	&0.273	&2.886	&41.290\\
				
				50&GeS&4.183/1.770&2.509	&35.321	&1.734	&32.968	&1.207	&7.671	&0.681	&2.824	&47.935\\
				
				51&Pb$_{2}$S$_{2}$&2.225/0.556&2.561	&4.5112	&0.65	&22.603	&0.452	&14.868	&0.151	&3.061	&71.616\\
				
				52&SnS&3.912/0.978&2.703	&33.105	&1.786	&22.529	&1.243	&14.919	&0.757	&2.795	&48.221\\
				
				53&SnSe&3.665/0.916&2.681	&24.671	&1.566	&29.150	&1.090	&10.322	&0.664	&2.794	&45.394\\
				
				54&Ge$_{2}$Se$_{2}$&1.861/0.465&2.542	&10.714	&0.509	&21.547	&0.354	&15.601	&0.167	&2.897	&60.238\\
				
				55&GeO&3.839/0.959&2.866	&43.654	&2.718	&60.142	&1.892	&11.197	&3.363	&2.455	&97.649\\
				
				56&S$_{2}$Si$_{2}$&2.884/0.721&2.750	&19.620	&1.504	&68.184	&1.047	&16.781	&1.13	&2.601	&25.975\\
				
				57&P$_{4}$&1.747/0.436&2.794	&51.555	&1.026	&14.399	&0.714	&20.564	&1.423	&2.424	&58.222\\
				
				58&As$_{4}$&1.563/0.390&2.725	&42.222	&0.759	&12.820	&0.528	&21.661	&0.77	&2.511	&14.074\\
				
			\end{tabular}
		}
	\end{ruledtabular}
	
	\footnotetext[1]{$G_{0}W_{0}$ band gap obtained from \cite{haastrup2018computational}},\footnotetext[2]{Jiang et al. \cite{PhysRevLett.118.266401} based binding energy calculation with E$_{\text{g}}/4$.}
	\footnotetext[3]{FCP model based $\beta$ tuned for intercepting E$_{\text{U}}$.}
	\footnotetext[4]{Relative MSE calculated with $\frac{\sqrt{(E_{\text{BSE}}-E_{\text{U}})^{2}}}{E_{\text{BSE}}} \times$100, Average Relative MSE = 25.954 $\%$ (see Figure \ref{figuresuup1}).}
	\footnotetext[5]{Calculated with 4$\times (13.6 \frac{\mu}{\epsilon^2})$ given by Wannier-Mott \cite{wannier,cheiwchanchamnangij2012quasiparticle} and intercepted by FCP model at $\beta_{\text{WM}} = 2.515$.}
	\footnotetext[6]{Relative MSE calculated with $\frac{\sqrt{(E_{\text{BSE}}-E_{\text{WM}})^{2}}}{E_{\text{BSE}}} \times$100, Average Relative MSE = 39.228 $\%$ (see Figure \ref{figuresuup1}).}
	\footnotetext[7]{FCP model calculation of binding energy with $\beta_{\text{mean}} = 2.625$.}
	\footnotetext[8]{Relative MSE calculated with $\frac{\sqrt{(E_{\text{BSE}}-E_{\text{FCPm}})^{2}}}{E_{\text{BSE}}} \times$100., Average Relative MSE = 12.809 $\%$ (see Figure \ref{figuresuup1}).}
	\footnotetext[8]{ Calculated with $\frac{8\mu}{(1+\sqrt{1+32\pi\alpha\mu/3})}$ given by Olsen et al. \cite{olsen2016simple}.}
	\footnotetext[9]{FCP model based $\beta$ tuned for intercepting E$_{\text{OLS}}$.}
	\footnotetext[10]{Relative MSE calculated with $\frac{\sqrt{(E_{\text{BSE}}-E_{\text{OLS}})^{2}}}{E_{\text{BSE}}} \times$100, Average Relative MSE = 52.980 $\%$ with outliers at 9$^{\text{th}}$ and 31$^{st}$ position not included (see Figure \ref{figuresuup1}).}
\end{table}
\begin{table}[h]
	\caption{\label{tab:table3}
		FCP model based calculation of experimentally reported binding energies for 8 bulk materials.}
	\begin{ruledtabular}
		\footnotesize
		\resizebox{0.8\columnwidth}{!}{
			\begin{tabular}{cccccccc}
				
				&Bulk&$\mu$&$\epsilon$&E$_{\text{rep}}\footnotemark[1]$(eV)&E$_{\text{FCP}}\footnotemark[2]$(eV)&$\beta_{\text{FCP}}\footnotemark[3]$&E$_{\text{A}}\footnotemark[4]$(eV)\\
				\hline
				1	&MoS$_{2}$	&0.42 \cite{fortin1975excitons}	&10.69 \cite{laturia2018dielectric}	&0.05 \cite{fortin1975excitons}	&0.05	&3	&0.052\\
				
				2	&MoSe$_{2}$	&0.62 \cite{anedda1980exciton}	&12.33 \cite{laturia2018dielectric}	&0.109 \cite{ugeda2014giant,anedda1980exciton}	&0.109	&2.99	&0.113\\
				
				3	&MoTe$_{2}$	&0.33 \cite{cao2013two}	&15.19 \cite{laturia2018dielectric}	&0.15 \cite{arora2017interlayer,beal1979kramers}	&0.150	&3.06	&0.020\\
				
				4	&WS$_{2}$	&0.23 \cite{beal1976excitons}	&7 \cite{beal1976excitons}	&0.064 \cite{beal1976excitons}	&0.063	&3	&0.066\\
				
				5	&WSe$_{2}$&0.21 \cite{beal1976excitons}	&7.3 \cite{beal1976excitons}	&0.055 \cite{habenicht2018mapping}	&0.053	&3	&0.056\\
				
				6	&1T-HfS$_{2}$	&3.75 \cite{habenicht2018mapping}	&7.7 \cite{habenicht2018mapping}	&0.17 \cite{habenicht2018mapping}	&0.17	&2.97	&0.141\\
				
				7	&h-BN	&0.54 \cite{cao2013two}	&4 \cite{cao2013two}	&$>$0.38 \cite{museur2011exciton,cao2013two}	&0.459	&3	&0.480\\
				
				9	&GaN	&0.17 \cite{shan1996binding}	&9.5 \cite{shan1996binding}	&0.021 \cite{shan1996binding}&0.022	&3	&0.023\\
				
				8	&P$_{4}$	&0.084 \cite{asahina1984band}	&11.29 \cite{asahina1984band}	&0.0079 \cite{asahina1984band}	&0.007	&3	&0.008\\
				
			\end{tabular}
		}
	\end{ruledtabular}
	\footnotetext[1]{Reported experimental binding energy.}
	\footnotetext[2]{Full solution based FCP calculation of experimental binding energy through tuning of $\beta$.}
	\footnotetext[3]{FCP model based $\beta$ tuned for intercepting E$_{\text{FCP}}$.}
	\footnotetext[4]{FCP based analytical calculation of binding energy using $\beta_{\text{FCP}}$.}
\end{table}
\end{document}